\documentstyle[12pt]{article}

\catcode`\@=11
\long\def\@makefntext#1{ 
\protect\noindent \hbox to 3.2pt {\hskip-.9pt
$^{{\ninerm\@thefnmark}}$\hfil}#1\hfill} 

\def\thefootnote{\fnsymbol{footnote}}
 \def\@makefnmark{\hbox to 0pt{$^{\@thefnmark}$\hss}}  

\def\ps@myheadings{\let\@mkboth\@gobbletwo
\def\@oddhead{\hbox{} 
\rightmark\hfil\ninerm\thepage}
\def\@oddfoot{}\def\@evenhead{\ninerm\thepage\hfil 
\leftmark\hbox{}}\def\@evenfoot{}
\def\sectionmark##1{}\def\subsectionmark##1{}}

\evensidemargin
0.30truein
\newcommand{\Dm}{D_{\mu}^+}
\newcommand{\Dbm}{D_{\mu}^-}
\newcommand{\Pm}{\partial_{\mu}^+}
\newcommand{\Pbm}{\partial_{\mu}^-}
\newcommand{\bibi}{\bibitem}

\newcommand{\gm}{\gamma}

\newcommand{\kp}{\kappa}

\newcommand{\ps}{\psi}

\newcommand{\Om}{\Omega}

\newcommand{\psb}{\overline{\ps}}

\newcommand{\hmu}{\hat{\mu}}

\newcommand{\RE}{\mbox{Re\,}}

\newcommand{\del}{\partial}

\newcommand{\dg}{\dagger}
\newcommand{\pr}{\prime}
\newcommand{\ra}{\rightarrow}

\newcommand{\be}{\begin{equation}}
\newcommand{\ee}{\end{equation}}
\newcommand{\bea}{\begin{eqnarray}}
\newcommand{\eea}{\end{eqnarray}}
\newcommand{\eq}{\ref}
\newcommand{\beq}{\begin{equation}}
\newcommand{\eeq}{\end{equation}}
\newcommand{\cc}{\cite}
\newcommand{\lb}{\label}

\def \3{\ss}

\begin{document}

\newcommand{\symbolfootnote}{\renewcommand{\thefootnote}
	{\fnsymbol{footnote}}}
\renewcommand{\thefootnote}{\fnsymbol{footnote}}
\newcommand{\alphfootnote}
	{\setcounter{footnote}{0}
	 \renewcommand{\thefootnote}{\sevenrm\alph{footnote}}}

\newcounter{sectionc}\newcounter{subsectionc}\newcounter{subsubsectionc}
\renewcommand{\section}[1] {\vspace{0.6cm}\addtocounter{sectionc}{1}
\setcounter{subsectionc}{0}\setcounter{subsubsectionc}{0}\noindent
	{\bf\thesectionc. #1}\par\vspace{0.4cm}}
\renewcommand{\subsection}[1] {\vspace{0.6cm}\addtocounter{subsectionc}{1}
	\setcounter{subsubsectionc}{0}\noindent
	{\it\thesectionc.\thesubsectionc. #1}\par\vspace{0.4cm}}
\renewcommand{\subsubsection}[1]
{\vspace{0.6cm}\addtocounter{subsubsectionc}{1}
	\noindent {\rm\thesectionc.\thesubsectionc.\thesubsubsectionc.
	#1}\par\vspace{0.4cm}}
\newcommand{\nonumsection}[1] {\vspace{0.6cm}\noindent{\bf #1}
	\par\vspace{0.4cm}}

\newcounter{appendixc}
\newcounter{subappendixc}[appendixc]
\newcounter{subsubappendixc}[subappendixc]
\renewcommand{\thesubappendixc}{\Alph{appendixc}.\arabic{subappendixc}}
\renewcommand{\thesubsubappendixc}
	{\Alph{appendixc}.\arabic{subappendixc}.\arabic{subsubappendixc}}

\renewcommand{\appendix}[1] {\vspace{0.6cm}
        \refstepcounter{appendixc}
        \setcounter{figure}{0}
        \setcounter{table}{0}
        \setcounter{equation}{0}
        \renewcommand{\thefigure}{\Alph{appendixc}.\arabic{figure}}
        \renewcommand{\thetable}{\Alph{appendixc}.\arabic{table}}
        \renewcommand{\theappendixc}{\Alph{appendixc}}
        \renewcommand{\theequation}{\Alph{appendixc}.\arabic{equation}}
        \noindent{\bf Appendix \theappendixc #1}\par\vspace{0.4cm}}
\newcommand{\subappendix}[1] {\vspace{0.6cm}
        \refstepcounter{subappendixc}
        \noindent{\bf Appendix \thesubappendixc. #1}\par\vspace{0.4cm}}
\newcommand{\subsubappendix}[1] {\vspace{0.6cm}
        \refstepcounter{subsubappendixc}
        \noindent{\it Appendix \thesubsubappendixc. #1}
	\par\vspace{0.4cm}}

\def\abstracts#1{{
	\centering{\begin{minipage}{30pc}\tenrm\baselineskip=12pt\noindent
	\centerline{\tenrm ABSTRACT}\vspace{0.3cm}
	\parindent=0pt #1
	\end{minipage} }\par}}

\newcommand{\bibit}{\it}
\newcommand{\bibbf}{\bf}
\renewenvironment{thebibliography}[1]
	{\begin{list}{\arabic{enumi}.}
	{\usecounter{enumi}\setlength{\parsep}{0pt}
\setlength{\leftmargin 1.25cm}{\rightmargin 0pt}
	 \setlength{\itemsep}{0pt} \settowidth
	{\labelwidth}{#1.}\sloppy}}{\end{list}}

\topsep=0in\parsep=0in\itemsep=0in
\parindent=1.5pc

\newcounter{itemlistc}
\newcounter{romanlistc}
\newcounter{alphlistc}
\newcounter{arabiclistc}
\newenvironment{itemlist}
    	{\setcounter{itemlistc}{0}
	 \begin{list}{$\bullet$}
	{\usecounter{itemlistc}
	 \setlength{\parsep}{0pt}
	 \setlength{\itemsep}{0pt}}}{\end{list}}

\newenvironment{romanlist}
	{\setcounter{romanlistc}{0}
	 \begin{list}{$($\roman{romanlistc}$)$}
	{\usecounter{romanlistc}
	 \setlength{\parsep}{0pt}
	 \setlength{\itemsep}{0pt}}}{\end{list}}

\newenvironment{alphlist}
	{\setcounter{alphlistc}{0}
	 \begin{list}{$($\alph{alphlistc}$)$}
	{\usecounter{alphlistc}
	 \setlength{\parsep}{0pt}
	 \setlength{\itemsep}{0pt}}}{\end{list}}

\newenvironment{arabiclist}
	{\setcounter{arabiclistc}{0}
	 \begin{list}{\arabic{arabiclistc}}
	{\usecounter{arabiclistc}
	 \setlength{\parsep}{0pt}
	 \setlength{\itemsep}{0pt}}}{\end{list}}

\newcommand{\fcaption}[1]{
        \refstepcounter{figure}
        \setbox\@tempboxa = \hbox{\tenrm Fig.~\thefigure. #1}
        \ifdim \wd\@tempboxa > 6in
           {\begin{center}
        \parbox{6in}{\tenrm\baselineskip=12pt Fig.~\thefigure. #1 }
            \end{center}}
        \else
             {\begin{center}
             {\tenrm Fig.~\thefigure. #1}
              \end{center}}
        \fi}

\newcommand{\tcaption}[1]{
        \refstepcounter{table}
        \setbox\@tempboxa = \hbox{\tenrm Table~\thetable. #1}
        \ifdim \wd\@tempboxa > 6in
           {\begin{center}
        \parbox{6in}{\tenrm\baselineskip=12pt Table~\thetable. #1 }
            \end{center}}
        \else
             {\begin{center}
             {\tenrm Table~\thetable. #1}
              \end{center}}
        \fi}

\def\@citex[#1]#2{\if@filesw\immediate\write\@auxout
	{\string\citation{#2}}\fi
\def\@citea{}\@cite{\@for\@citeb:=#2\do
	{\@citea\def\@citea{,}\@ifundefined
	{b@\@citeb}{{\bf ?}\@warning
	{Citation `\@citeb' on page \thepage \space undefined}}
	{\csname b@\@citeb\endcsname}}}{#1}}

\newif\if@cghi
\def\cite{\@cghitrue\@ifnextchar [{\@tempswatrue
	\@citex}{\@tempswafalse\@citex[]}}
\def\citelow{\@cghifalse\@ifnextchar [{\@tempswatrue
	\@citex}{\@tempswafalse\@citex[]}}
\def\@cite#1#2{{$\null^{#1}$\if@tempswa\typeout
	{IJCGA warning: optional citation argument
	ignored: `#2'} \fi}}
\newcommand{\citeup}{\cite}

\def\fnm#1{$^{\mbox{\scriptsize #1}}$}
\def\fnt#1#2{\footnotetext{\kern-.3em
	{$^{\mbox{\sevenrm #1}}$}{#2}}}

\font\twelvebf=cmbx10 scaled\magstep 1
\font\twelverm=cmr10 scaled\magstep 1
\font\twelveit=cmti10 scaled\magstep 1
\font\elevenbfit=cmbxti10 scaled\magstephalf
\font\elevenbf=cmbx10 scaled\magstephalf
\font\elevenrm=cmr10 scaled\magstephalf
\font\elevenit=cmti10 scaled\magstephalf
\font\bfit=cmbxti10
\font\tenbf=cmbx10
\font\tenrm=cmr10
\font\tenit=cmti10
\font\ninebf=cmbx9
\font\ninerm=cmr9
\font\nineit=cmti9
\font\eightbf=cmbx8
\font\eightrm=cmr8
\font\eightit=cmti8


\centerline{\tenbf CHIRAL GAUGE THEORIES ON THE LATTICE WITHOUT GAUGE FIXING?}
\baselineskip=22pt
\centerline{\tenrm WOLFGANG BOCK}
\baselineskip=13pt
\centerline{\tenit Department of Physics, University of California, San
Diego,}
\baselineskip=12pt
\centerline{\tenit La Jolla, CA 92093-0319, USA}
\vspace{0.9cm}
\abstracts{We discuss two proposals for  a non-perturbative
formulation of chiral gauge theories on the lattice. In both
cases gauge symmetry is broken by the regularization. We aim at a
dynamical restoration of symmetry. If the gauge symmetry breaking
is not too severe this procedure could lead in the continuum limit
to the desired chiral gauge theory.
}

\vfil
\twelverm   
\baselineskip=14pt
\noindent {\bf 1. Restoration of gauge symmetry}.
The non-perturbative formulation of chiral gauge theories
on the lattice is still an unsolved problem in field theory.
On a hypercubic
lattice each Weyl fermion is accompanied by 15 species
doublers, where half of these carry opposite chirality.
A naive lattice transcription of a chiral gauge theory leads therefore
to a vector-like theory, instead of the desired chiral gauge
theory. Many of the methods, which have been proposed to solve this
problem, conflict with
the concept of gauge invariance. We shall discuss here two methods,
Wilson's approach and the staggered fermion approach, which have
been used before in vector-like theories to the reduce the number of
fermion flavors. The first method tries to
remove the species doublers from the spectrum
by rendering them heavier than the
cut-off, whereas the second one
uses them as physical degrees of freedom. Both methods lead to
gauge non-invariant actions when applied to chiral gauge theories.
One possibility
to treat these non-gauge invariant models is to start from a
gauge fixed continuum model and to adopt one of the two methods to
transcibe it to the lattice \cc{ROME}. Since  the resulting
lattice action breaks  BRST invariance, one has to
add all counterterms with dimension $\leq 4$ and tune their
coefficients such that this symmetry gets restored in the scaling region.
{}From a technical point of view however
this method is very cumbersome and one also has to worry about
non-perturbative gauge fixing.
Alternatively one can also aim at  a {\em dynamical restoration
of gauge invariance}. Both            models become automatically
gauge invariant after integrating over the gauge fields
in the path intergal, however at the price of introducing an extra
radially frozen scalar field $V_x$ \cc{Jan}.
Let's start first from a generic non-gauge invariant lattice action,
$S(U)$. It is easy to show
that the partition function can be written in the following form
$Z = \int D U e^{  S(U_{\mu x})}
= \int D U D V e^{ S(V_x^{\dg} U_{\mu x} V_{x+\hmu})}$.
The new form of the action $S(V_x^{\dg} U_{\mu x} V_{x+\hmu})$
is  trivially invariant under the local gauge transformations
$U_{\mu x} \ra \Om_x U_{\mu x} \Om^{\dagger}_{x+\hmu}$,
$V_x \ra \Om_x V_x$. The important question of symmetry restoration
is whether the resulting model
still can describe the physics of the underlying gauge invariant target
model in the continuum.  \\

\noindent {\bf 2. Wilson's method}.
As our target model
we shall consider here a U(1)$_L^{\rm{local}} \otimes $  U(1)$_R^{\rm{global}}$
gauge-fermion model. Its lattice action is given by the following expression
\be
S=- \frac{1}{2} \sum_{x \mu} \psb^\pr_x \gm_{\mu}
  \left[(\Dm+\Dbm)P_L + (\Pm+\Pbm)P_R \right] \psi_x^\pr
  -y \sum_x \psb_x^\pr  \psi_x^\pr
  +\frac{w}{2} \sum_x  \psb_x^\pr  \Box
   \psi^\pr_x  \lb{WIL} \;,
\ee
where $\Dm  \psi^\pr_x \equiv  U_{\mu x} \psi^\pr_{x+\hmu} -\psi^\pr_x$,
$\Dbm \psi^\pr_x \equiv  \psi^\pr_x - U_{\mu x-\hmu}^* \psi^\pr_{x-\hmu}$,
$\del^{\pm}_{\mu} =D^{\pm}_{\mu}|_{U=1}$,
$\Box \equiv \del_{\mu}^+ \del_{\mu}^-$
and a standard Wilson mass term has been added to
remove the species doublers from the spectrum. Both, the Wilson mass-and
the bare mass term break chiral gauge invariance.
The action which results after integrating over the
gauge fields in the partition function is given by (\eq{WIL}) with
$U_{\mu x} \ra  V_x^* U_{\mu x} V_{x+\hmu}$.
The $\psi^\pr_L$-field in this action is  screened from
the gauge fields by the $V$-fields and
is  therefore {\em neutral} with respect to
the U(1)$_L$ gauge transformations. By the gauge transformation
$\psi^\pr_x  = (V_x^* P_L+ P_R) \psi_x$
we can remove the $V$-fields from the kinetic term.
The action in terms of the $\psi$-fields
is identical with (\eq{WIL}) (with $\psi^\pr \ra \psi$), except
that the bare-and Wilson mass terms turn into Yukawa-and
Wilson-Yukawa terms,
$-y \sum_x ( \psb_{Lx} V_x \psi_{Rx} + \psb_{Rx} V_x^* \psi_{Lx})$ and
$+\frac{w}{2} \sum_x ( \psb_{Lx} V_x \Box \psi_{Rx} +
\psb_{Rx} \Box ( V_x^{*} \psi_{Lx} ))$.
We have studied this model, which is known  also as Smit-Swift model
\cc{SMIT},
in the global symmetry limit, $U_{\mu x}=1$, and with
the term $2\kp \sum_{x \mu} \RE ( V_x^*  V_{x+\hmu})$
added to the action. The phase diagram which has
been established  by numerical and analytical calculations
is shown in fig.~1. Besides the ferromagnetic (FM) phase there
are two symmetric phases, PMS and PMW. Analytic and numerical calculations
have shown that the physics in the
PMW (PMS) phase is described by the action in terms of the
$\psi$ ($\psi^\pr$)-fields. The particles which are
associated with the scalar fields decouple
in both phases as long as one keeps far away from the
phase boundaries.
The $\psi$-fermion in the PMW phase is massless. However, its species
doublers stay also massless and the resulting model is vector-like.
In contrast in the PMS phase the species doublers of the $\psi^\pr$-fermion
can indeed be removed from the spectrum,  but
the physical $\psi^\pr$-fermion
is massive and decouples from the bosonic particles in the
continuum limit \cc{DON}. It has also be shown
that a charged fermion, which would couple to
gauge fields and may exist as a
$V$-$\psi^\pr$ bound state, does not exist in the PMS phase \cc{DON}.
This result shows that a dynamical restoration of chiral gauge symmetry does
nowhere occur in the phase diagram.
The same model exhibits however also an example for dynamical gauge
symmetry restoration: Let's consider for the
moment the vector-like naive fermion model with $w=y=0$ as our traget model.
The gauge invariance of this model gets broken for $y>0$.
The $\kp$-$y$ phase diagram is given by the
$w=0$-plane in fig.~1. The $\psi$-fermion
is  massless in the PMW phase, even though the coefficient $y$ of the
symmetry breaking mass term is non-zero and gauge symmetry is restored in
this phase provided that
one keeps far away   from the phase boundaries.
%
%
\begin{figure}[t]
\vspace*{5.0cm}
\caption{ \noindent {\em Phase diagram of the Smit-Swift model.
}}
\label{FIG1}
\end{figure}
\\
\noindent {\bf 3. The staggered fermion approach}.
Staggered fermions describe
four Dirac flavors in the scaling region.
The Dirac and flavor components
of these four staggered flavors do not appear in an explicit form
since they are spread out over the lattice.
It has been shown in ref.~\cc{Jan} how  one can couple these
spin and flavor components in an arbitrary manner to other fields.
The staggered fermion action may  symbolically
be represented by the first term in (\eq{WIL}), keeping in mind
however that the spin-flavor components of the $\psi^\pr$ field
are now spread out over the lattice.
The model lacks gauge invariance since
gauge transformations on the $U$-fields cannot
be carried through to the $\psi^\pr$-fields.
A perturbative calculation in two dimensions
has shown that the model performs well
for smooth external gauge fields \cc{SMOOTH}.
The important issue however is whether this remains true also when
taking into account the full quantum fluctuations.
The gauge invariant version of the staggered fermion
model is again obtained by
$U_{\mu x} \ra  V_x^* U_{\mu x} V_{x+\hmu}$.
Since the staggered fields $\psi^\pr$ are neutral
with respect to the U(1) transformations we have added also the
mass term in (\eq{WIL}) to the action.
In contrast to the Wilson-Yukawa model we cannot
relate the $\psi^\pr$-fields by a gauge
transformation to a ``staggered'' $\psi$-field.
The crucial question
however is whether the scalar fields are sufficiently
smooth such that this transformation
could effectively take place, leading to a PMW
phase at small $y$ with a massless $\psi$-fermion.
In this case gauge
symmetry would be restored dynamically.
An extensive investigation of the $\kp$-$y$ phase
diagram for the case of an axial-vector model (this model
has a larger lattice symmetry group) has shown that
the PMS phase extends down to $y=0$ with no
PMW phase opening up at small $y$ \cc{ST4} and thereto also in this case
the desired restoration of gauge symmetry does not take place. \\
\noindent {\bf 4. Acknowledgements}: This work was done  in collaboration with
A.K. De, J. Smit and J. Vink.
The research was supported by the ``Stichting voor
Fun\-da\-men\-teel On\-der\-zoek der Materie (FOM)'',
by the ``Stichting Nationale Computer Faciliteiten (NCF)'' and by
the DOE under contract DE-FG03-91ER40546. \\
\noindent {\bf 5. References:}

\end{document}